\documentclass[aps,prl,twocolumn,showpacs,superscriptaddress]{revtex4}
\usepackage{graphicx}
\usepackage{amsmath,amssymb,latexsym}
\bibstyle{apsrev.bib}

\begin{document}

\title{Ranking knots of random, globular polymer rings}

\author{M. Baiesi}
\affiliation{Dipartimento di Fisica, Universit\`a di Firenze,
and Sezione INFN, Firenze,
I-50019 Sesto Fiorentino, Italy.} 
\author{E. Orlandini}
\affiliation{Dipartimento di Fisica and Sezione CNR-INFM,
Universit\`a di Padova, I-35131 Padova, Italy.} 
\affiliation{Sezione INFN, Universit\`a di
Padova, I-35131 Padova, Italy.}
\author{A. L. Stella}
\affiliation{Dipartimento di Fisica and Sezione CNR-INFM, 
Universit\`a di Padova, I-35131 Padova, Italy.}
\affiliation{Sezione INFN, Universit\`a di
Padova, I-35131 Padova, Italy.}

\begin{abstract}

An analysis of extensive simulations of interacting self-avoiding 
polygons on cubic lattice shows that the frequencies of different 
knots realized in a random, collapsed polymer ring decrease as a
negative power of the ranking order, and suggests that the total number of 
different knots realized grows exponentially with the chain length.
Relative frequencies of specific knots converge to definite values because 
the free energy per monomer, and its leading finite size corrections, 
do not depend on the ring topology, while a
subleading correction only depends on the crossing number of the knots.
\end{abstract}


\maketitle

Issues related to the probability of realization of configurations 
with specific knots in closed random chains play a major role in 
topological polymer statistics~\cite{GENERAL1,DeWittStu88} 
and in its applications to macromolecular 
and biological physics~\cite{GENERAL2}. 
Interest in the spectrum of different knots realized 
in random polymer models is stimulated, e.g., by the need of comparison with 
the circular DNA extracted from some viral capsids~\cite{capsids,Micheletti}, 
in the hope to 
identify specific biological mechanisms of knot formation. 
The low number of globular proteins for which a knot has been detected in 
the native state~\cite{Taylor,protein_knots}
marks a striking difference with respect to the general 
collapsed phase of homopolymers, in which
a definitely higher knotting frequency is expected. 
Understanding the 
reasons of such a difference is certainly a key issue for the formulation 
of adequate statistical models of proteins~\cite{protein_knots_2}. 
The interest in collapsed polymers is also stimulated by the recent 
realization that their knots, unlike prime knots in the good solvent 
case, are on average completely delocalized along the backbone~\cite{pd1}. 
Understanding if and up
to what extent topological invariants can affect the globular
state in such conditions is an intriguing fundamental issue. 
Already in the swollen regime, the problem of precisely
determining the possible dependence on topology of the free
energy per monomer and of the exponent specifying its  
correction $\propto \ln N / N$ in the limit where the 
number $N$ of monomers approaches infinity, remains 
open~\cite{entropy_swollen,preparation}. 
This in spite of the fact that the localized character of prime knots 
should represent a simplifying feature. In the globular state similar
issues have never been addressed and their discussion
should include the free energy
correction associated to the existence of the globule-solvent 
interface~\cite{Owczarek,Baiesi06}. 

In the present Letter we investigate the different topologies realized 
by the equilibrium configurations of a collapsed ring polymer. 
In spite of the considerable complexity of these configurations,
we show that a relatively simple statistical law governs the frequencies of 
the various knots, with far reaching consequences. 
On one side, it allows to argue the rate at which the 
amplitude of the spectrum of different knot topologies grows with 
increasing ring length.
At the same time, the emerging scenario confirms that knots are
delocalized and clarifies how topology controls
the statistics of the globular state.

A cornerstone in topological polymer statistics has been the realization 
that, for self-avoiding polygons (SAP's), 
unknotted configurations are entropically 
disfavored, so that their probability approaches zero exponentially with 
growing chain length~\cite{unknot_N_0,unknot_N_0_b,preparation}. 
However, after this important step, progress in the 
statistical analysis of knot complexity was hindered by the circumstance 
that unknotted configurations in swollen, good solvent regimes remain
overwhelmingly dominant even for relatively very long chains.
On the other hand, for polymer rings in bad solvent, the collapsed 
state leads to expect a much higher probability of knotting compared to 
the swollen case with the same chain length. This circumstance
suggests collapsed polymers as ideal systems for the study of how 
topological complexity develops and statistically
distributes itself with growing chain length.

As a model of the large scale behavior of a long flexible polymer
in a solvent we adopt the self-avoiding walk on cubic lattice.
Attractive energies ($\varepsilon=-1$)
between non-consecutive nearest neighbor visited 
sites allow us to work in the collapsed state at $T < T_{\Theta}$,
where $T_{\Theta}$ is the theta-collapse temperature~\cite{Vanderzande}.
In such regime it is rather difficult to sample a sufficient
number of uncorrelated polymer configurations. We use the pruned 
enriched Rosenbluth method~\cite{PERM} (PERM) which revealed very effective 
and was successfully used also for the search of native states of 
protein models~\cite{nPERM}. Our computational
problem is made heavier by the fact that the closed chain 
configurations, i.e.~the SAP's generated 
by PERM are only a small fraction of the total. A more serious 
difficulty is the topological analysis of the knot type present in a given
closed chain configuration.
The configurations of long, collapsed 
SAP's are very intricate geometrically and their 
projections on planes present a huge number of crossings. This makes 
impossible the calculation of the topological invariants necessary
for the knot identification. In order to circumvent this
difficulty, we simplify each sampled configuration
before performing the analysis of invariants. To this purpose we
apply to the configuration a smoothing algorithm, which progressively
reduces the length of the chain, 
while keeping its topology unaltered (for a similar procedure, 
see~\cite{Micheletti}).
This algorithm is in fact a BFACF grand-canonical simulation~\cite{BFACF}
in which the step fugacity is fixed low enough to cause
a rapid reduction of step number in the SAP 
to be analyzed. Projections of such shrunk SAP are then analyzed by 
the HOMFLY polynomial in ``Knotscape'' program~\cite{Knotscape}. 
Our code allows us to resolve prime knots up to 11 essential crossings, 
and composite knots up to 5 components.

\begin{figure}[!tb] 
\includegraphics[angle=0,width=7.6cm]{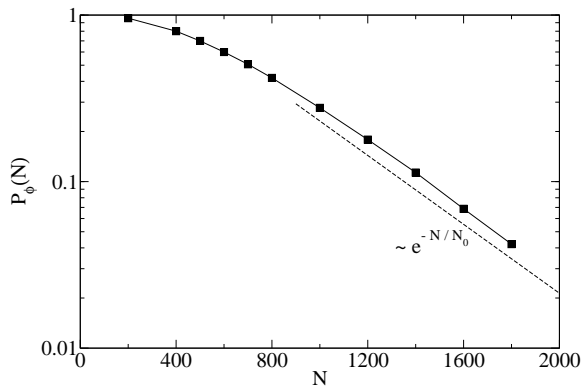}
\caption{Probability of finding an unknot. The 
straight dashed line represents the exponential fit with $N_0=420$.
\label{fig:1}} 
\end{figure} 

Along the above lines we perform a systematic analysis
of the knot spectrum for $N$-step SAP's at
$T = 2.5 < T_{\Theta}$ ($T_{\Theta}= 3.595$~\cite{theta}), for 
chain lengths up to $N=1800$. 
A first result concerns the behavior of the probability  $P_{\emptyset}(N)$
of configurations with knot $k=\emptyset$, where
$\emptyset$ indicates the unknot,  as a function of $N$. 
Our procedure allows the classification of almost
all configurations for $N$ up to $800$, while from $N=1000$ to $N=1800$ we have
a progressive degradation of performances, reaching a
$59\%$ of unresolved configuration for $N=1800$. There is thus an uncertainty
on the normalization needed to calculate probabilities 
like $P_{\emptyset}(N)$. 
The most plausible scenario is that unresolved 
configurations have complex knots,
especially prime knots with more than 11 crossings or composite
ones with more than 5 prime components.
They enter thus in the statistics as ``unresolved''.
We see an exponential decay $P_{\emptyset} \sim e^{- N / N_0 }$, 
for increasing $N$. The decay constant is $N_0 \simeq 420$
(see Fig.~\ref{fig:1}). 
This value is two orders of magnitude smaller than that expected for 
$T=\infty$ ($N_0\approx 2.4\times 10^5$~\cite{unknot_N_0,unknot_N_0_b}), 
when SAP statistics is controlled by excluded
volume alone. This small $N_0$ indicates that configurations with
knots have appreciable probability already for relatively short
SAP's. 
In a recent work on Hamiltonian loops on the cubic lattice~\cite{Grosberg04}
a value of $N_0\lesssim 196$ was estimated, 
about half the one we determine at $T=2.5$.
If we think of Hamiltonian loops as a ($T=0$)-like situation, 
a lower value of $N_0$ should indeed be expected if the trend 
of $N_0$ decreasing with $T$ is general.

\begin{figure}[!tb]  
\includegraphics[angle=0,width=7.6cm]{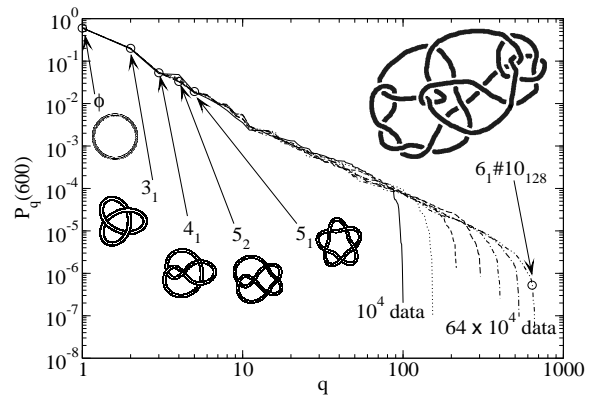}
\caption{
Probability of rank-ordered knots for N=600.
Curves are for sample sizes increasing as a power of $2$, from $10^4$ to
 $64 \times 10^4$. Some simple knot topologies 
(unknot, $3_1$, $4_1$, $5_2$, and $5_1$) 
and a more complicated one ($6_1\#10_{128}$)
are also associated with their ranking.
\label{fig:cutoff}} 
\end{figure}

A further step was the analysis of the knot type $k$ of each sampled
configuration and of its probability $P_k(N)$ in the statistics. 
We rank in decreasing order these probabilities and, by simply 
indicating as $P_q$ the probability of the knot $k$ with rank $q$
($q=1, 2, 3, \ldots$),
we obtain the log-log plots reported in Fig.~\ref{fig:cutoff}.
These plots correspond to increasing numbers of sampled configurations
for $N=600$.
All curves display the same slope and overlap in the first part.
They only differ for the cutoff: the richer the sample, the
larger the maximum rank of the realized knots. 
At the same time, upon varying sample size, we observe 
stability of the rank ordering of the knots corresponding to
the initial linear parts of the plots. 
Thus, the estimates of $P_k(600)$ should not
be affected by systematic errors.
All this means that the observed cutoff in rank is only due 
to limited sampling and has nothing to do with the presumably 
much higher cutoff on the spectrum of different realizable knots due to 
the fact that the SAP length is finite.
The power law behavior shown by all the plots is therefore a robust 
feature of the data in this regime.

\begin{figure}[!tb] 
\includegraphics[angle=0,width=7.6cm]{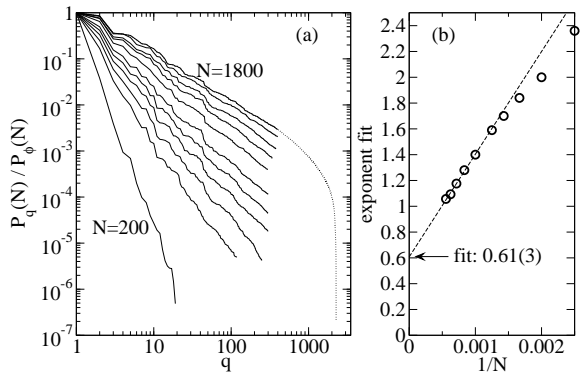}
\caption{(a)
Power-law ranges of $P_q(N)/P_\emptyset(N)$,
for $N=200$, $400$, $500$, $600$, $700$,
$800$, $1000$, $1200$, $1400$, $1600$, and $1800$.
For $N=1800$ also the cutoff, due to limited sampling, is shown.
(b) Extrapolation of the exponent of the Zipf law for $N\to\infty$.
\label{fig:3}} 
\end{figure}

Figure~\ref{fig:3} shows similar plots, this time for varying $N$
and with quite rich samples. For clarity the effects of
the statistical cutoffs are not shown, and probabilities are divided
by the unknot probability $P_{\emptyset}(N)$.
The slope of the plots is 
an increasing function of $N$. The estimates of the slope 
extrapolate to a value $-0.61(3)$ (see Fig.~3) for
$N \to \infty$. Thus, for large $N$
the rank ordering statistics of the knots obeys a law of the
Zipf type:
\begin{equation}
P_{q}(\infty)/P_{\emptyset}(\infty) \sim q^{-r}
\label{eq:1}
\end{equation}
with $r \approx 0.6$.
The Zipf law was first observed in
the context of linguistics~\cite{Zipf}, where it rules the rank
ordering in frequency of different words in a text. Since then it
emerged in different fields, ranging from economics to
disordered systems~\cite{Zipf-2}. For sure the validity of such
simple law for knots in random polymers
reveals a remarkable and unsuspected degree of organization 
of such form of topological complexity. 

The asymptotic validity of Eq.~(\ref{eq:1}) with exponent $r < 1$
allows us to infer a fundamental property of the
knot spectrum, which at first sight would seem 
unaccessible to any numerical investigation.
As mentioned above, there must be a cutoff $q_{\rm max}(N)$
for the Zipf law~(\ref{eq:1}).
Even if the statistical cutoff due to finite sampling
occurs at much lower $q$'s, the dependence of $q_{\rm max}(N)$ 
on $N$ is a key information on the spectrum, providing a lower bound for
the maximum number of different 
knot topologies a SAP of length $N$ can host. 
Assuming validity of the Zipf behavior in Eq.~(\ref{eq:1})
up to $q=q_{\rm max}(N)$, in view of the normalization condition
$\sum_q P_q(N)=1$ valid for any $N$, one easily
concludes that
\[
1+\sum_{q=2}^{q_{\rm max}(N)} q^{-r} 
\sim q_{\rm max}^{1-r} \sim P_{\emptyset}(N)^{-1}.
\]
Thus, $q_{\rm max}(N) \sim P_{\emptyset}(N)^{{-1}/({1-r})} \sim e^{N/0.4 N_0}$.
This means that the number of different possible topologies a SAP
can host grows (at least) exponentially with $N$.

The validity of the law in Eq.~(\ref{eq:1}) also for the asymptotic  
ranking of knots is induced by some balance in the knot frequencies, 
which should not diverge relative to each other for $N\to\infty$.
For a collapsed globule in equilibrium, theoretical arguments 
and numerical results suggest the following form for the 
large $N$ behavior of the canonical partition function~\cite{Owczarek}:
\begin{equation}
e^{-F/kT} \sim A\, e^{\kappa N} e^{\kappa_1 N^{2/3}} N^{\alpha-2}
\label{eq:2}
\end{equation}
where $F$ is the free energy, $\kappa$ and $\kappa_1$ are
reduced free energies per monomer and per interface
monomer, respectively, and $\alpha$ is an 
exponent~\cite{Vanderzande}.
The stretched exponential factor containing $\kappa_1$
clearly implies an interfacial contribution to the
free energy, since the area of the globule-solvent 
interface is expected to grow $\sim N^{2/3}$.
It is natural to expect an asymptotic behavior
of the form in Eq.~(\ref{eq:2}) also for ensembles with
fixed knot topology like those considered here.
The value of $\alpha$
could depend on the topology of the globule in such ensembles.
Indeed, in the $T \to \infty$ case, there are 
indications that the analog of $\alpha$
for prime knots differs sensibly from that for composite 
knots~\cite{entropy_swollen} (as confirmed in~\cite{preparation}). 
Dependences on topology could not be
excluded, a priori, also for parameters like $\kappa$ and $\kappa_1$.
Indeed, the equivalent of $\kappa$ in the $T \to \infty$ limit
for the unknotted ring is rigorously known to be different 
from the $\kappa$ 
in the ensemble with unrestricted topology~\cite{DeWittStu88}.
On the other hand, the $\kappa$ of the unknotted ring has been
conjectured to be the same as that of any other knotted
ring for non-interacting SAP's~\cite{entropy_swollen}.  

\begin{figure}[!tb] 
\includegraphics[angle=0,width=7.6cm]{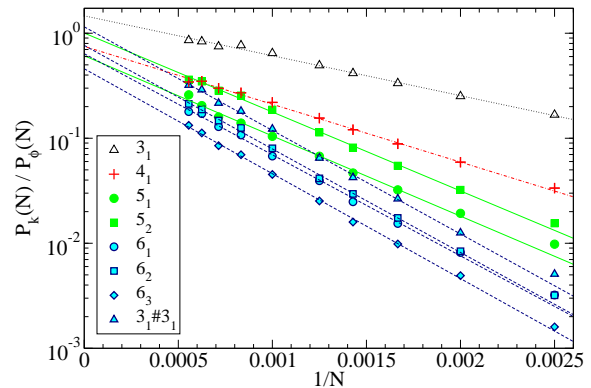}
\caption{(Color online) 
Probability of a knot type $k$ over the probability of an unknot, 
vs $1/N$, for some knots. Each ratio converges 
towards an asymptotic value $A_k / A_{\emptyset}$.
Straight lines are fits $\sim \exp[ - (\delta_k-\delta_{\emptyset}) / N]$.
\label{fig:4}} 
\end{figure} 

\begin{figure}[!tb] 
\includegraphics[angle=0,width=7.6cm]{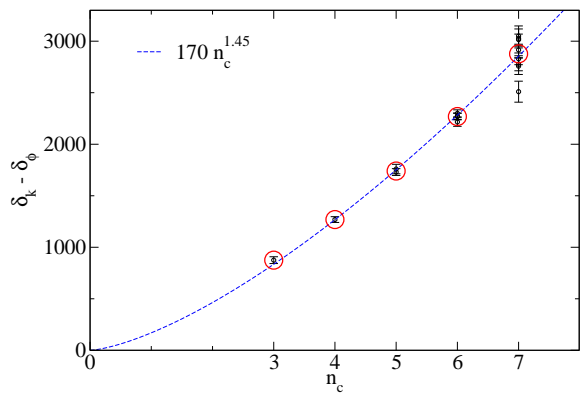}
\caption{(Color online) Fitted $\delta_k-\delta_{\emptyset}$ 
as a function of the minimal crossing number $n_c$ for knots 
$3_1$ ($n_c=3$), $4_1$ ($n_c=4$), $5_1$ and $5_2$ ($n_c=5$), $3_1\#3_1$, 
$6_1$, $6_2$, and $6_3$ ($n_c=6$), and for all knots with $n_c=7$.
(Large circles denote averages for each $n_c$). 
$\delta_k$'s are grouped in bands corresponding to $n_c$'s, i.e. to a good 
approximation they depend only on the number of essential crossings of a knot
type.
A fit is  also shown, suggesting that $\delta_k - \delta_{\emptyset}$ 
grows as a power of $n_c$. 
\label{fig:5}} 
\end{figure} 

An analysis of ratios $P_k/P_{\emptyset}$ should reveal the 
possible dependence of $\kappa$, $\kappa_1$, $\alpha$, and $A$
on the knot type. 
These ratios are shown in 
Fig.~\ref{fig:4} as a function of $1/N$ for the simplest prime and composite
knots~\cite{note1}. 
They do not diverge. This implies that for the analyzed different knots
in the collapsed polymer ring the parameters $\kappa$, $\kappa_1$, and $\alpha$
must be the same. In particular, unlike in the
$T \to \infty$ case, the entropic $\alpha$ exponent should
be the same for all knots. 
On the other hand, Fig.~\ref{fig:4} reveals that one should include
a knot-dependent subleading factor $\exp(- \delta_k / N)$ in the
form of the partition function (\ref{eq:2}). 
The differences $\delta_k - \delta_{\emptyset}$
are proportional to 
the slopes of data sets in log-scale in Fig.~\ref{fig:4}. These have
values that essentially are determined by the crossing number $n_c$ of the
knots (see Fig.~\ref{fig:5}). 
For example, $\delta_{3_1\#3_1} \approx \delta_{6_1}$, and so on.
A simple power-law increase  $\delta_k - \delta_{\emptyset}\sim n_c^{1.45}$
fits rather well the data in Fig.~\ref{fig:5}. 

We stress that for swollen polymers 
there are clear numerical evidences~\cite{entropy_swollen, preparation} 
of the conjecture that
$\alpha_k = \alpha_{\emptyset} + \pi_k$, where $\pi_k$ is the number of 
prime components of the knot. This can be explained by taking into account 
that each such component is localized along the 
chain~\cite{pd1,entropy_swollen, preparation},
bringing an entropic factor $\sim N$ to the partition function.
On the other hand, the convergence of the relative frequencies of knots 
to definite values rules out this 
picture for collapsed polymers 
($\alpha_k = \alpha_{\emptyset}$).

In summary, we have shown that a great deal of information
can be gained from a numerical investigation of knots in 
collapsed polymers. The study of ratios of knot frequencies and 
the established Zipf type of law for
the ranking in frequency of the knots, allow to
draw very solid conclusions concerning the universality
with respect to topology of several statistical
parameters characterizing the globules. 
We find that the 
only non-universal parts of the knot frequencies relative to a reference 
frequency (the one of the unknotted rings) are an 
asymptotic ($N\to \infty$) knot-dependent amplitude ratio and the rate of
convergence to this ratio, which seems to be determined only by 
the crossing number of the knots. 
These results are consistent with the expectation that 
knots are delocalized in the collapsed regime~\cite{pd1}.
The Zipf law also enables us to predict how the spectrum of different
topologies grows in amplitude with growing ring length. 

We thank A.~Stasiak for discussions.
This work was supported by FIRB01 and  MIUR-PRIN05.
M.B. acknowledges financial support from EC FP6 project 
``EMBIO''
(EC contract nr.~012835).


\begin{thebibliography}{99}
\bibitem{note1}
Note that asymptotically $3_1$ becomes slightly more frequent
than $\emptyset$. The unknot however remains the most frequent knot 
if topologies are distinguished also according to their chiralities, 
such that the two mirror versions $3_1^+$ and $3_1^-$ represent two
different knots.
\end{thebibliography}
\end{document}